\documentclass[10pt,letterpaper,twocolumn]{article}
\usepackage{ol2}
\usepackage{varioref}
\usepackage{graphicx}
\usepackage[T1]{fontenc}
\usepackage{subfigure}
\usepackage{cite}
\usepackage{amsmath}
\usepackage{amsfonts}
\usepackage{stmaryrd}
\usepackage{psfrag}
\usepackage{tabularx}\newcolumntype{Y}{>{\centering\arraybackslash}X}
\graphicspath{{/home/image/eps/pmma/}}
\DeclareGraphicsExtensions{.eps,.EPS,.ps,.PS,.eps.gz}
\begin{document}

\twocolumn[
\title{Unidirectional sub-diffraction waveguiding based on optical spin-orbit coupling in subwavelength plasmonic waveguides}

\author{Yannick Lefier and Thierry Grosjean}

\address{Universit\'e  de Franche-Comt\'e, Département d'Optique P.M. Duffieux,\\ Institut
FEMTO-ST, UMR CNRS 6174\\ 15B Av. des Montboucons, 25030 Besançon cedex, France}

\email{thierry.grosjean@univ-fcomte.fr}

\begin{abstract}

Subwavelength plasmonic waveguides show the unique ability of strongly localizing (down to the nanoscale) and guiding light. These structures are intrinsically two-way optical communication channels, providing two opposite light propagation directions. As a consequence, when light is coupled to these planar integrated devices directly from the top (or bottom) surface using strongly focused beams, it is equally shared into the two opposite propagation directions. Here, we show that symmetry can be broken by using incident circularly polarized light, on the basis of a spin-orbital angular momentum transfer directly within waveguide bends.  We predict that up to 94 \% of the incoupled light is directed into a single propagation channel of a gap plasmon waveguide. Unidirectional propagation of strongly localized optical energy, far beyond the diffraction limit, becomes switchable by polarization, with no need of intermediate nano-antennas/scatterers as light directors. This study may open new perspectives in a large panel of scientific domains, such as nanophotonic circuitry, routing and sorting, optical nanosensing, nano-optical trapping and manipulation.
\end{abstract}
\maketitle ]


Subwavelength plasmonic waveguiding has drawn a considerable interest during the past years for îts unique ability of controlling light down to the nanometer scale, opening the perspective of highly integrated optical circuits and ultra-compact optical functions \cite{ebbesen:phystod08}. Several plasmon waveguide geometries, such as metallic V-grooves \cite{gramotnev:apl04,bozhevolnyi:nature06}, nanostripes \cite{charbonneau:ol00}, nanowires \cite{dickson:jpcb00,ditlbacher:prl05}, nanogaps \cite{pile:apl05,veronis:ol05,liu:ox05}, wedges \cite{pile:ol05,moreno:prl08}, dielectric-loaded metal films \cite{holmgaard:ox09} have been proposed for strongly confining and guiding light. Given their intrinsic symmetry, plasmonic waveguides provide two-way propagation channels of opposite directions. Generally, light is coupled into the waveguide mode with end-firing techniques in order to reach unidirectional propagation of light at subwavelength scale. This coupling technique avoids one of the two possible propagation directions: propagation reversal within the waveguide requires two different coupling devices positioned at its two extremities.

Recently, reversible unidirectional light propagation has been obtained onto planar metallic surfaces (with surface plasmons) \cite{rodriguez-fortuno:sci13,lin:sci13,mueller:nl14}, in photonic crystal waveguides \cite{kapitanova:natcom14}, in nanofibers  \cite{petersen:sci14,mitsch:natcom14} and in dielectric stripes \cite{rodriguez:acsphot14}. All these studies are based on the coupling of angular momentum between a rotating dipolar nano-emitter and the evanescent surface waves involved in the waveguiding process, on the basis of spin-orbit interaction in localized fields \cite{onoda:prl04}: the intrinsic chirality of the evanescent waves in play makes the connection between the point-like emitter and the waveguide\cite{bliokh:pra12,bliokh:natcomm14}. This technique allows for reversing the propagation direction of the waveguide mode by switching circular polarization direction. Reversible unidirectional guiding has also been achieved in dielectric waveguides with incident linear polarization \cite{rodriguez:ol14}. For all these techniques however, light coupling into the waveguide needs to be mediated by a nanoscale scatterer placed in contact to the waveguide, which may represent challenging fabrication processes. Moreover, the waveguides in play are often limited by diffraction and do not possess the confinement ability of plasmonic waveguides.

Here, we propose a different approach to reach reversible unidirectional light waveguiding while breaking down the diffraction barrier. Our concept is based on spin-orbital angular momentum coupling within the curvature of a polarization-sensitive subwavelength under excitation with a circularly polarized out-of-plane focused beam. This configuration, which avoids the need of any external point-like scatterers or nano-antenna to couple light into the waveguide, leads to highly versatile nano-optical platforms.


Let us consider a circularly polarized plane wave propagating along $z$-direction. The expression of its complex electric-field amplitude ($\vec{E}$) in cylindrical coordinates $(r,\phi,z)$ takes the following from:

\begin{eqnarray}
     \vec{E} &=& E_r \vec{e}_r + E_{\phi} \vec{e}_{\phi}\\
     \vec{E} &\propto& (\vec{e}_r \pm i \vec{e}_{\phi})\exp(\pm i\phi)
\end{eqnarray}

where $\vec{e_r}$ and $\vec{e_{\phi}}$ are unit vectors along radial and azimuthal directions, respectively. Term $\exp(\pm i\phi)$ in Eq. 1 defines a spiral phase for both the radial and azimuthal field components. A circularly polarized plane wave can thus be seen as the superposition of two radially and azimuthally polarized vortices described by the same topological charge $\pm 1$ and in phase quadrature. When this circularly polarized plane wave is coupled to a polarization sensitive structure that interacts selectively with $E_r$ or $E_{\phi}$ and avoids the other field component in its intrinsic optical process, circular polarization can be converted into a single radially or azimuthally polarized vortex by the structure. The photon spin (carried by circular polarization) is then transferred into an orbital angular momentum (manifested by field vorticity), leading to a circulating optical energy flow. Such a spin-orbit interaction of light has been observed for example in the excitation of axis-symmetrical surface plasmon waves with circularly polarized focused beams \cite{cho:ox12}. In that case, axis-symmetrical "p"-polarized surface plasmons are selectively excited from the radially polarized vortex component of the incident circular polarization (no interaction with the azimuthally polarized vortex $E_{\phi}$), leading to surface plasmon vortices.

According to the above explained optical process, we propose to generate optical spin-orbit interaction in subwavelength plasmon waveguides, which form a unique family of polarization sensitive light conveyors, far beyond the diffraction limit. To show the intended effect, we have chosen the gap plasmon waveguide whose nanometer scale confinement and guiding properties are bound to a capacitive effect in-between two facing metallic edges \cite{pile:apl05,veronis:ol05,liu:ox05}. This capacitive effect originates polarization sensitivity properties of the waveguide. Note that the other plasmonic waveguides evoked in the introduction could also be considered.

\begin{figure}[htbp]
\begin{center}
\includegraphics [width=0.5\columnwidth]{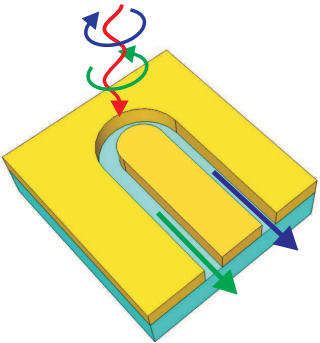}
\caption{Scheme of the proposed concept of unidirectional subwavelength optical waveguiding with a subwavelength plasmonic waveguide.}\label{fig:scheme}
\end{center}
\end{figure}

The principle of the proposed unidirectional mode coupling is schemed in Fig. \ref{fig:scheme}. Circularly polarized light is directly focused (along the out-of-plane $z$-axis) onto the local bend of a plasmonic waveguide. When light is focused onto a straight gap plasmon waveguide, the guided plasmon mode is excited selectively from the field component perpendicular to the pair of facing edges, due to transverse optical capacitive effect with respect to waveguide propagation axis. Note that such a configuration also leads to surface plasmon launching but relatively large slits are used for that purpose, for which the optical capacitive effect is weak \cite{cho:ox12}. Narrower slits, which dramatically increase the optical capacitive effect and thus improve subwavelength waveguiding, are less efficient in surface plasmons launching \cite{lalanne:prl05}. Given their radial shape and polarization sensitivity, bent subwavelength plasmonic waveguides placed within a circularly polarized focused beam  will be selectively coupled (by scattering) with just one of the two axially polarized vortices which constitute circular polarization (see Eq. 2). For example, a bent gap plasmon waveguide is selectively coupled to the radially polarized vortex of the circularly polarized focused beam. The guided mode excited at the waveguide bend then undergoes spin-induced orbital angular momentum whose orientation is fully determined by the circular polarization direction (i.e. the spin orientation). Therefore, depending on the direction of the incident circular polarization, a circulating energy flow along the waveguide path may occur "upstream" or "downstream", leading to polarization-controlled unidirectionally in the excitation of deeply subwavelength guided modes.

\begin{figure}[htbp]
\begin{center}
\includegraphics [width=0.8\columnwidth]{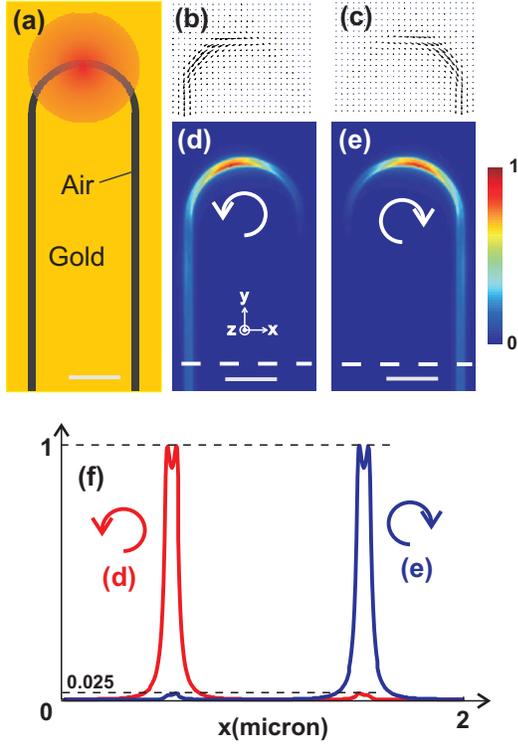}
\caption{Unidirectional waveguiding in a "U"-shape subwavelength gap plasmon waveguide with circularly polarized focused beam. (a) scheme of the waveguide structure. Focus width and position are shown with a red disk (out-of-plane focusing with a 0.8 numerical aperture objective is considered). (b) and (c): 2D vector maps of the poynting vector in a transverse plane (xy) located 10 nm far from the metal surface. (d) and (e): electric intensity plots along the same transverse plane for (d) left and (e) right circular polarization. Scale bars: 500 nm. (f): electric intensity profile plotted along the dashed lines of (d) and (e), for the two opposite circular polarization directions.}\label{fig:U}
\end{center}
\end{figure}
\begin{figure}[htbp]
\begin{center}
\includegraphics [width=0.8\columnwidth]{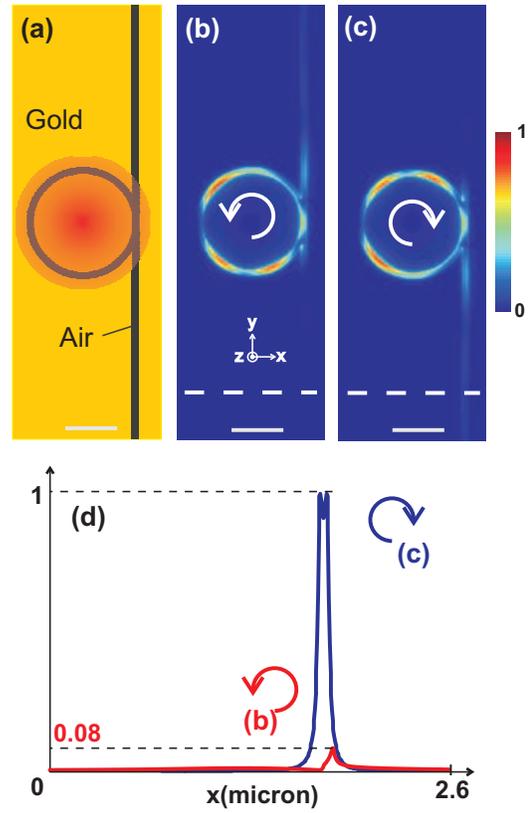}
\caption{Unidirectional waveguiding at a loop of gap plasmon waveguide with circularly polarized focused beam. (a) scheme of the waveguide structure. Focus width and position are shown with a red disk (out-of-plane focusing with a 0.8 numerical aperture objective is considered). (b) and (c): electric intensity distribution along a transverse (xy)-plane located 10 nm far from the metal surface,  for (b) left and (c) right circular polarization. Scale bars: 500 nm. (d): electric intensity profile plotted along the dashed lines of (b) and (c), for the two opposite circular polarization direction.}\label{fig:loop}
\end{center}
\end{figure}

This concept is validated numerically with Finite Difference Time Domain (FDTD) method available from the commercial code "`Fullwave"' (Synopsis). Two types of waveguide curvatures are under study. First, we consider the half-circle curvature, giving to the waveguide a "U"-shape (Fig. \ref{fig:U}(a)), and second, we investigate the total circle curvature, i.e. the waveguide loop (Fig. \ref{fig:loop}(a)). In both cases, the gap plasmon waveguide is engraved in a 50 nm thick gold layer lying onto a dielectric substrate of refractive index equal to 1.5. The gap plasmon waveguide is 50 nm wide and the radius of curvature of both waveguide bends is 0.5 $\mu$m, in order to be compatible with focusing systems of modest numerical apertures. The wavelength is 1550 nm in all calculations. The computation volume is $6.4\mu m \times 3 \mu m \times 1.8 \mu m$ and all its six boundaries are terminated with Perfectly Matched Layers in order to avoid parasitic unphysical reflections around the structure. In all simulations, non-uniform grid resolution varies from 20 nm for portions at the periphery of the simulation, to 5 nm in the region including the waveguide bend. The waveguides are excited with a gaussian beam propagating along the out-of-plane axis $(0z)$. The beam waist is 1.2 micron wide (1/e full width) and it is placed directly under the waveguide bend, in the substrate. We model here light focusing with a 0.8 numerical aperture objective. The position of the focused beam with respect to the subwavelength plasmonic waveguide is shown with a red circular area in Figs. \ref{fig:U}(a) and \ref{fig:loop}(a).

Figures \ref{fig:U}(d) and (e) show the distribution of electric optical intensity along the transverse (xy)-plane  (parallel to metal surface) located 10 nm beyond the waveguide, in air, with incident right and left circular polarization, respectively. We clearly see that light is channeled along the left arm of the U-shape waveguide (right arm, respectively) for incident left circular polarization (right circular, respectively).  The channeled plasmon stays strongly localized right at the waveguide, well beyond the excitation area, proving nanometer scale waveguiding. In Fig. \ref{fig:U}(f) are plotted the electric intensity profiles along the white dashed lines of Fig. \ref{fig:U}(d) and (e). The intensity maximum at the excited arm is about 40 times larger than at the other arm, for both right and left circular polarizations. Note that light scattering at the waveguide bend can launch surface plasmons, leading to spurious optical effects. However, we see in Fig. \ref{fig:U}(d) and (e) that intensity level of the surface plasmons is much smaller than the intensity of the guided mode and no coupling can occur between surface plasmons and the guided mode because of their effective index mismatch \cite{veronis:ol05,liu:ox05}. Figures \ref{fig:U}(b) and (c) display 2D vectorial maps of the poynting vector distribution along the same (xy)-plane (10 nm far from the metal surface), for left and right circular polarizations, respectively. These maps are plotted in the region of the waveguide bend. We clearly see an azimuthal flow of poynting vector, right at the bent gap plasmon waveguide, that is reversible by switching circular polarization direction. As noted previously, the spin-orbit optical interaction at the waveguide bend, combined to the waveguide polarization sensitivity, couples the radially polarized vortex component of the incident circularly polarized beam into the waveguide mode, leading to a circulating poynting flow within the waveguide bend and thus, unidirectional waveguiding properties despite symmetric structure and illumination. The guided optical power ($P$) in each arm of the U-shape waveguide can be written as $P=\iint_{S} \vec{\pi} \cdot \vec{dS}$ where $\vec{\pi}$ is the poynting vector and $S$ is the area of a transverse cross-section of the waveguide.
For a given circular polarization direction, the fraction $\rho$ of the incoupled optical power that is channeled within the excited arm is defined  as $\rho=P_{exc}/(P_{exc}+P_{non-exc})$ where $P_{exc}$ and $P_{non-exc}$ are the guided powers along the excited and non-excited arms, respectively. Calculation of the guided power across 100-nm wide transverse cross-sections, located 2 wavelengths away from the excitation zone (i.e. 2 wavelengths away from the waveguide bend), show that 93.8 \% of the incoupled light is guided into a single propagation direction.

Figure \ref{fig:loop} shows the optical properties of the waveguide loop under excitation with a circularly polarized focused beam.  As for the U-shape gap plasmon waveguide, we clearly see the unidirectional waveguiding process that occurs along the straight portion of the plasmonic waveguide. Left  circular polarization leads to waveguiding along $+y$ direction (Fig. \ref{fig:loop}(b)) whereas right circular polarization transfers energy along the opposite $-y$ direction (Fig. \ref{fig:loop}(c)). Figure \ref{fig:loop}(d) displays intensity profiles along the dashed line shown in \ref{fig:loop}(b) and (c). The intensity maximum at the excited arm is about 12.5 times higher than the intensity at the other arm. Calculation of the poynting vector flow across 100-nm wide transverse waveguide cross-sections of both channel arms show that 93.3 \% of the incoupled light is guided into a single propagation direction. As for the U-shape gap plasmon waveguide, the spin-orbital angular momentum transfer generates a circulating power flow in the waveguide loop (whose left or right direction is defined by circular polarization) that induces unidirectional launching of confined light within the straight portion of plasmonic waveguide.


To conclude, we propose the concept of unidirectional nanoscale waveguiding that is switchable by the helicity of a circularly polarized excitation beam. Unidirectionality is induced by a spin-orbit angular momentum interaction directly within waveguide bends under local excitation with circularly polarized focused light. No nanoscale antennas or scatterers are needed for directing the energy within the waveguide. The resulting optical architectures are thus versatile. Improved mode coupling efficiency may however be achieved by considering more sophisticated launchers such as gratings, but at the expense of more complex devices. Note that spin-orbit conversion is predictable in the bends of all polarization-sensitive waveguides, i.e. subdiffraction plasmonic waveguide but also dielectric waveguides operating at the microscale. This study may open new routes in a large panel of scientific domains, such as nanophotonic circuitry, routing and sorting, optical nanosensing, nano-optical trapping and manipulation. \\


This work is supported by the Labex ACTION.


\begin{thebibliography}{10}
\newcommand{\enquote}[1]{``#1''}
\expandafter\ifx\csname url\endcsname\relax
  \def\url#1{\texttt{#1}}\fi
\expandafter\ifx\csname urlprefix\endcsname\relax\def\urlprefix{URL }\fi
\providecommand{\eprint}[2][]{\url{#2}}

\bibitem{ebbesen:phystod08}
T.~Ebbesen, C.~Genet, and S.~Bozhevolnyi, \enquote{Surface-plasmon circuitry,}
  Physics Today \textbf{61}, 44 (2008).

\bibitem{gramotnev:apl04}
D.~K. Gramotnev and D.~F.~P. Pile, \enquote{Single-mode subwavelength waveguide
  with channel plasmon-polaritons in triangular grooves on a metal surface,}
  Applied Physics Letters \textbf{85}, 6323 (2004).

\bibitem{bozhevolnyi:nature06}
S.~Bozhevolnyi, V.~Volkov, E.~Devaux, J.~Laluet, and T.~Ebbesen,
  \enquote{Channel plasmon subwavelength waveguide components including
  interferometers and ring resonators,} Nature \textbf{440}, 508 (2006).

\bibitem{charbonneau:ol00}
R.~Charbonneau, P.~Berini, E.~Berolo, and E.~Lisicka-Shrzek,
  \enquote{Experimental observation of plasmon polariton waves supported by a
  thin metal film of finite width,} Opt. Lett. \textbf{25}, 844
  (2000).

\bibitem{dickson:jpcb00}
R.~Dickson and L.~Lyon, \enquote{Unidirectional Plasmon Propagation in Metallic
  Nanowires,} J. Phys. Chem. B \textbf{104}, 6095 (2000).

\bibitem{ditlbacher:prl05}
H.~Ditlbacher, A.~Hohenau, D.~Wagner, U.~Kreibig, M.~Rogers, F.~Hofer, F.~R.
  Aussenegg, and J.~R. Krenn, \enquote{Silver Nanowires as Surface Plasmon
  Resonators,} Phys. Rev. Lett. \textbf{95}, 257403 (2005).

\bibitem{pile:apl05}
D.~F.~P. Pile, T.~Ogawa, D.~K. Gramotnev, T.~Okamoto, M.~Haraguchi, M.~Fukui,
  and S.~Matsuo, \enquote{Theoretical and experimental investigation of
  strongly localized plasmons on triangular metal wedges for subwavelength
  waveguiding,} Applied Physics Letters \textbf{87}, 061106 (2005).

\bibitem{veronis:ol05}
G.~Veronis and S.~Fan, \enquote{Guided subwavelength plasmonic mode supported
  by a slot in a thin metal film,} Opt. Lett. \textbf{30}, 3359 (2005).

\bibitem{liu:ox05}
L.~Liu, Z.~Han, and S.~He, \enquote{Novel surface plasmon waveguide for high
  integration,} Opt. Express \textbf{13}, 6645(2005).

\bibitem{pile:ol05}
D.~F.~P. Pile and D.~K. Gramotnev, \enquote{Plasmonic subwavelength waveguides:
  next to zero losses at sharp bends,} Opt. Lett. \textbf{30}, 1186
  (2005).

\bibitem{moreno:prl08}
E.~Moreno, S.~G. Rodrigo, S.~I. Bozhevolnyi, L.~Martin-Moreno, and F.~J.
  Garcia-Vidal, \enquote{Guiding and Focusing of Electromagnetic Fields with
  Wedge Plasmon Polaritons,} Phys. Rev. Lett. \textbf{100}, 023901 (2008).

\bibitem{holmgaard:ox09}
T.~Holmgaard, Z.~Chen, S.~I. Bozhevolnyi, L.~Markey, and A.~Dereux,
  \enquote{Dielectric-loaded plasmonic waveguide-ring resonators,} Opt. Express
  \textbf{17}, 2968 (2009).

\bibitem{rodriguez-fortuno:sci13}
F.~Rodriguez-Fortuno, G.~Marino, P.~Ginzburg, D.~O'Connor, A.~Martinez,
  G.~Wurtz, and A.~Zayats, \enquote{Near-field interference for the
  unidirectional excitation of electromagnetic guided modes,} Science
  \textbf{340}, 328 (2013).

\bibitem{lin:sci13}
J.~Lin, J.~B. Mueller, Q.~Wang, G.~Yuan, N.~Antoniou, X.-C. Yuan, and
  F.~Capasso, \enquote{Polarization-controlled tunable directional coupling of
  surface plasmon polaritons,} Science \textbf{340}, 331 (2013).

\bibitem{mueller:nl14}
B.~Mueller, K.~Leosson, and F.~Capasso, \enquote{Polarization-Selective
  Coupling to Long-Range Surface Plasmon Polariton Waveguides,} Nano lett.
  \textbf{14}, 5524 (2014).

\bibitem{kapitanova:natcom14}
P.~V. Kapitanova, P.~Ginzburg, F.~J. Rodrìguez-Fortuño, D.~S. Filonov, P.~M.
  Voroshilov, P.~A. Belov, A.~N. Poddubny, Y.~S. Kivshar, G.~A. Wurtz, and
  A.~V. Zayats, \enquote{Photonic spin Hall effect in hyperbolic metamaterials
  for polarization-controlled routing of subwavelength modes,} Nat. commun.
  \textbf{5}, 3326 (2014).

\bibitem{petersen:sci14}
J.~Petersen, J.~Volz, and A.~Rauschenbeutel, \enquote{Chiral nanophotonic
  waveguide interface based on spin-orbit interaction of light,} Science
  \textbf{346}, 67 (2014).

\bibitem{mitsch:natcom14}
R.~Mitsch, C.~Sayrin, B.~Albrecht, P.~Schneeweiss, and A.~Rauschenbeutel,
  \enquote{Quantum state-controlled directional spontaneous emission of photons
  into a nanophotonic waveguide,} Nat. commun. \textbf{5}, 5713 (2014).

\bibitem{rodriguez:acsphot14}
F.~J. Rodriguez-Fortuno, I.~Barber-Sanz, D.~Puerto, A.~Griol, and A.~Martinez,
\enquote{Resolving light handedness with an on-chip silicon microdisk,} ACS
Photonics \textbf{1}, 762 (2014).

\bibitem{onoda:prl04}
M.~Onoda, S.~Murakami, and N.~Nagaosa, \enquote{Hall Effect of Light,} Phys.
  Rev. Lett. \textbf{93}, 083901 (2004).

\bibitem{bliokh:pra12}
K.~Y. Bliokh and F.~Nori, \enquote{Transverse spin of a surface polariton,}
  Phys. Rev. A \textbf{85}, 061801 (2012).

\bibitem{bliokh:natcomm14}
K.~Y. Bliokh, A.~Y. Bekshaev, and F.~Nori, \enquote{Extraordinary momentum and
  spin in evanescent waves,} Nat. commun. \textbf{5}, 3300 (2014).

\bibitem{rodriguez:ol14}
F.~J. Rodriguez-Fortuno, D.~Puerto, A.~Griol, L.~Bellieres, J.~Marti, and
A.~Martinez, \enquote{Sorting linearly polarized photons with a single
scatterer,} Opt. Lett. \textbf{39}, 1394 (2014).

\bibitem{cho:ox12}
S.-W. Cho, J.~Park, S.-Y. Lee, H.~Kim, and B.~Lee, \enquote{Coupling of spin
  and angular momentum of light in plasmonic vortex,} Opt. Express
  \textbf{20}, 10083 (2012).

\bibitem{lalanne:prl05}
P.~Lalanne, J.-P. Hugonin, and J.-C. Rodier, \enquote{Theory of surface plasmon
  generation at nanoslit apertures,} Phys. Rev. Lett. \textbf{95}, 263902
  (2005).

\end{thebibliography}

\end{document}